\documentclass[aps,prd,superscriptaddress,showkeys,showpacs,onecolumn,nofootinbib,notitlepage,10pt]{revtex4-2}
\usepackage{amsmath,amssymb,slashed,color}
\usepackage{graphicx}
\usepackage{hyperref}
\usepackage{bm}
\usepackage{comment}

\DeclareMathOperator{\Ai}{Ai}
\def\bal#1\gal{\begin{align}#1\end{align}}

\newcommand{\eq}[1]{(\ref{#1})}
\newcommand{\fig}[1]{Fig.~\ref{#1}}
\renewcommand{\sec}[1]{Sec.~\ref{#1}}

\begin{document}

\title{Quasi-Classical Approximation of Electromagnetic Radiation by Fermions embedded in Rigidly Rotating Medium in Strong Magnetic Field}

\author{Matteo Buzzegoli}\email{matteo.buzzegoli@e-uvt.ro}
\affiliation{Department of Physics, West University of Timișoara, Bd. Vasile Pârvan 4, Timișoara 300223, Romania}

\affiliation{
Department of Physics and Astronomy, Iowa State University, Ames, Iowa, 50011, USA}

\author{Kirill Tuchin}\email{tuchink@gmail.com}

\author{Nandagopal Vijayakumar}\email{nvmg@iastate.edu}

\affiliation{
Department of Physics and Astronomy, Iowa State University, Ames, Iowa, 50011, USA}

\begin{abstract}

We develop the quasi-classical (WKB) approximation of the synchrotron radiation by a fermion embedded into uniformly rotating system in external magnetic field. 
We show that it gives an accurate approximation of the exact expression that we recently obtained at a tiny fraction of the numerical cost. Our results can be used to compute the electromagnetic radiation of the quark-gluon plasma produced in relativistic heavy-ion collisions.

\end{abstract}

\maketitle

\section{Introduction}\label{sec:intro}

We have recently developed a theory of synchrotron radiation by fermions embedded into rotating systems \cite{Buzzegoli:2022dhw,Buzzegoli:2023vne,Buzzegoli:2023yut,Buzzegoli:2024nzd} with the aim of understanding the impact of rotation on the electromagnetic radiation by the quark-gluon plasma, which was recently observed to posses high vorticity collinear with the magnetic field \cite{Csernai:2013bqa,Csernai:2014ywa,Becattini:2015ska,Deng:2016gyh,Jiang:2016woz,Kolomeitsev:2018svb,Deng:2020ygd,Xia:2018tes}. We derived analytical results for the intensity of the electromagnetic radiation and used them to numerically investigate its dependence on the angular velocity of rotation. We showed that depending on the relative direction of the angular velocity and the magnetic field, the radiation intensity can be enhanced or suppressed. The main focus of  
\cite{Buzzegoli:2022dhw,Buzzegoli:2023vne} was on the systems rotating with `relatively slow' angular velocities satisfying $\Omega^2\ll |qB|$. We argued that the boundary conditions on the light-cylinder have little impact on the radiation intensity. In faster-rotating systems, the boundary conditions on the light-cylinder become essential and were discussed in \cite{Buzzegoli:2023yut,Buzzegoli:2024nzd}.

In principle, one may use the results obtained in \cite{Buzzegoli:2022dhw,Buzzegoli:2023vne,Buzzegoli:2023yut,Buzzegoli:2024nzd} to compute the electromagnetic radiation by the quark-gluon plasma. The synchrotron radiation of non-rotating electromagnetic plasmas was studied in \cite{Baring:1988a,Herold:1982a,Harding:1987a}. However, such a calculation in rotating plasmas turns out to be very expensive numerically because rotation lifts the degeneracy of the spectrum, which not only makes analytical expressions way more complicated than in the non-rotating case but also greatly increases the complexity of the numerical calculations. It is, therefore, important to develop a reliable approximation. In this paper, we develop the quasi-classical approximation, which hinges on the assumption that spacing between the nearby Landau levels is negligible. Considering that the distance between the nearby Landau levels equals the synchrotron frequency $\omega_B$, the applicability of the quasi-classical approximation requires that $\omega_B\ll E$, where $E$ is the fermion energy. It can be argued that  \cite{berestetskii1982quantum}
\begin{align}\label{eq:est}
\frac{\omega_B}{E} \sim \frac{B}{M^2/q}\frac{M^2}{E^2}= \frac{qB}{E^2}\,,
\end{align}
where $M$ is the fermion mass.
This indicates that the quasi-classical approximation is expected to be accurate in describing the synchrotron radiation of ultra-relativistic fermions in subcritical magnetic fields. This argument applies as well to a rotating system where $E$ is understood as including the additional term $-m\Omega$, where $m$ is the magnetic quantum number:
\begin{equation} \label{eq:energy}
    E= \sqrt{M^2 + p_z^2 + 2 n |q B|} + m\Omega\,,
\end{equation}
where $p_z$ is the conserved fermion momentum along the magnetic field, which we choose to be the $z$-axis. Evidently, rotation lifts the degeneracy of the Landau levels in $m$. It was argued in \cite{Buzzegoli:2024nzd} (see also \cite{Buzzegoli:2022omv}) that if $\Omega^2\sim qB$, the structure of the Landau levels is completely broken due to the lifted degeneracy in $m$ so that it is no longer possible to give a simple estimate of the distance between the levels. We therefore constrain ourselves in this paper to the `relatively slow' rotation. The development of the quasi-classical approximation in such a case is the main subject of this paper.

In our analysis, we follow the method Sokolov and Ternov
\cite{book:SokolovAndTernov,book:Bordovitsyn}
who first obtained the expressions in the leading order perturbation theory and then expanded it in the quasi-classical limit. It is a natural method in our case since we have already derived the leading order result and we proceed along the same guidelines. 

Our paper is organized as follows. In \sec{sec:exact}, we review the baseline results obtained in \cite{Buzzegoli:2022dhw,Buzzegoli:2023vne}. The radiation intensity involves the Laguerre function $I_{n,n'}$ that parametrizes the matrix element of photon emission. The WKB approximation is developed in \sec{sec:wkb} and used in \sec{sec:quasiclassics} to obtain the quasi-classical expression for the radiation intensity. \sec{sec:results} is devoted to numerical analysis. We summarize and conclude in \sec{sec:summary}. 

Throughout the paper, $q$ denotes the electric charge carried by the fermion, and $\bm B$ and $\bm \Omega$ are the collinear magnetic field and the angular velocity of rotation.  We adopt the natural units $\hbar=c=1$ unless otherwise indicated, and the quantities shown in the figures are given in units of the fermion mass, $M=1$.

\section{Synchrotron Radiation Intensity}\label{sec:exact}

It is instructive to begin with the review of the main result of \cite{Buzzegoli:2022dhw,Buzzegoli:2023vne}. The Landau levels of a particle in a constant magnetic field embedded in a uniformly rotating medium are given by 
\eq{eq:energy}. Since the operator of rotation about the $z$-axis commutes with the boost operator in the same direction, we can choose a reference frame where the momentum of  the initial electron along the $z$-axis vanishes: $p_z=0$.
When $\bm\Omega$ is parallel to $\bm B$, $\Omega$ is taken to be positive and negative when it is anti-parallel. In place of $m$, it is often useful to use the radial quantum number $a$: 
\begin{equation} \label{eq:radial quantum no}
    a = n - m - \frac{1}{2}\,.
\end{equation}
The quantum numbers $n$ and $a$ takes integer values from $0$ to $\infty$. We will use the prime such as $E'$, $a'$, etc.\  to distinguish the final state. 
The radiation intensity  is given by 
\begin{equation} \label{eq:intensity_formula}
    W^{h}_{n,a,p_z=0,\zeta} = \frac{q^2}{4 \pi}\sum_{n',a',\zeta '} \int \omega^2 d\omega \int d\cos \theta \delta(\omega-E+E^{\prime})  \Phi_h\,,
\end{equation}
where $\zeta, \zeta^{\prime}$ denotes the initial and final fermion polarization, $h$ denotes the helicity of the emitted photon, and $\theta$ denotes the emission angle of the photon with respect to the $z$-axis. The delta function imposes energy conservation, and momentum conservation is implicit. $\Phi_h$ is the matrix element of the scattering process given by
\begin{align}\label{eq:matrix_element}
    \Phi_h = I^2_{a,a'}(x)\big|\sin \theta \left( K_4 I_{n-1,n'-1}(x) - K_3 I_{n,n'}(x) \right) + K_1 (h - \cos \theta) I_{n,n'-1} (x)\nonumber\\
    + K_2 (h + \cos \theta) I_{n-1,n'} (x)\big|^2\,.
\end{align}
The function $I_{n,n'}(x)$ is defined as 
\begin{equation} \label{eq:I-func_def}
    I_{n,n'}(x) = \sqrt{\frac{n^{\prime}!}{n!}} e^{-x/2} x^{(n-n')/2} L^{n-n'}_{n'}(x)\,,
\end{equation}
where $L^{n-n'}_{n'}(x)$ is the associated Laguerre polynomial and 
\begin{align} \label{eq:x}
    x = \frac{\omega^2 \sin^2\theta}{2|qB|}\,.
\end{align}
 $I_{n,n'}(x)$ is sometimes called the  Laguerre function. We will call these functions the $I$-functions throughout the article. 
 The coefficients $K_i$ that appears in the intensity are defined as follows:
\begin{subequations} \label{eq:Ks}
    \begin{align}
        K_1 &= C_1'C_4 + C_3'C_2\,,\qquad 
        K_2 = C_4'C_1 + C_2'C_3\,, \label{eq:Ks1}\\
        K_3 &= C_4'C_2 + C_2'C_4\,,\qquad 
        K_4 = C_1'C_3 + C_3'C_1\,, \label{eq:Ks2}
    \end{align}
\end{subequations}
 where
 \begin{align}\label{eq:Cs}
 C_{1,3} = \frac{1}{2\sqrt{2}} B_3 (A_3 \pm A_4)\,,\qquad
 C_{2,4} = \frac{1}{2\sqrt{2}} B_4 (A_4 \mp A_3),
 \end{align}
 with the upper signs refer to the indexes 1 and 2, and the lower ones to 3 and 4.
 The primed quantities are defined similarly with primed $A$ and $B$ coefficients, which are as follows: 
\begin{subequations} \label{eq:coeff-2}
    \begin{align}  \label{eq:As}
        A_3 &= 1\,,\qquad
        A_4  = \zeta \,,  \qquad
        B_3 = \left( 1 +\frac{\zeta M}{ E - m\Omega}\right)^{1/2}\,,\qquad
        B_4 = \zeta \left( 1 -\frac{\zeta M}{ E - m\Omega}\right)^{1/2}\,,  
    \end{align}
    \begin{align} \label{eq:prime_coeff}
        A'_3 &= \left( 1 -\frac{\omega \cos{\theta}}{E' - m'\Omega}\right)^{1/2}\qquad \, \quad ,\,\qquad
        A'_4 = \zeta'\left( 1 +\frac{\omega \cos{\theta}}{E' - m'\Omega}\right)^{1/2}\,,  \\
        B'_3 &= \left( 1 +\frac{\zeta' M}{\sqrt{ (E' - m'\Omega)^{ 2}-p_z'^2}}\right)^{1/2}\,,\quad
        B'_4 = \zeta'\left( 1 -\frac{\zeta' M}{\sqrt{(E' - m'\Omega)^{ 2}-p_z'^2}}\right)^{1/2}\, .  
    \end{align}
 \end{subequations}

The total intensity of photons produced from a given initial energy (associated with the quantum numbers, $n$ and $a$) of the fermion is obtained by summing over the photon helicities, $h$, and averaging over the initial fermion polarizations, i.e., $\zeta$ in \eq{eq:intensity_formula} which gives us the following result:
\begin{align} \label{eq:matrix_elt_helcty_sum}
      \Phi = \sum_{h =\pm 1} \Phi_h  = 2 I^2_{a,a'}(x) \Big\{\left[(K_4 I_{n-1,n'-1}-K_3I_{n,n'})\sin \theta - (K_1 I_{n,n'-1}
     +K_2I_{n-1,n'})\cos \theta\right]^2\nonumber\\ 
      +(K_1 I_{n,n'-1}-K_2I_{n-1,n'})^2\Big\} \,,
 \end{align}
\begin{align} \label{eq:W_expr_1}
    W_{n,a} = \frac{1}{2}\sum_{\zeta,h}W^{h}_{n,a,p_z=0,\zeta} = \frac{q^2}{4 \pi}\sum_{n',a'} \int \omega^2 d\omega \int d(\cos \theta) \delta(\omega-E+E^{\prime})\left(\frac{1}{2} \sum_{\zeta,\zeta^{\prime}} \Phi\right)\,.
\end{align}
The integral over $\omega$ can be performed by integrating the delta-function (as done in \cite{buzzegoli2023photon}.) To do so, we require the solution to the equation,
\begin{align} \label{eq:Energy_conv}
    \omega = E - E^{\prime}\,.
\end{align}
Conservation of momentum along the longitudinal direction dictates that
\begin{equation} \label{eq:momentum_conv}
    p_z^{\prime} = p_z - \omega \cos \theta\,.
\end{equation}
Using \eq{eq:energy} and \eq{eq:momentum_conv}, we can solve for $\omega$ in \eq{eq:Energy_conv}, and the solution is as follows:
\begin{equation} \label{eq:omega_sol}
    \omega_0 = \frac{E-m'\Omega-p_z\cos\theta}{\sin^2\theta} \left\{1-\left[1-\frac{\mathcal{B}\sin^2\theta}{(E-m'\Omega-p_z\cos\theta)^2}\right]^{1/2}\right\}\,,
\end{equation}
where
\begin{equation}
    \mathcal{B} = 2(n-n^{\prime})|qB| - (m-m^{\prime})^2\Omega^2 + 2(m-m^{\prime})\Omega(E-m^{\prime}\Omega)\,.
\end{equation}
Since we perform our computations in a frame where the initial longitudinal momentum vanishes ($p_z = 0$), we have
\begin{equation} \label{eq:momentum_conv_0}
    p_z^{\prime} =  - \omega \cos \theta \, ,
\end{equation}
\begin{equation} \label{eq:omega_naught}
     \omega_0 = \frac{E-m'\Omega}{\sin^2\theta} \left\{1-\left[1-\frac{\mathcal{B}\sin^2\theta}{(E-m'\Omega)^2}\right]^{1/2}\right\}.
\end{equation}
The delta-function then becomes
\begin{equation} \label{eq:delta_rewrite}
    \delta(\omega-E+E^{\prime}) = \frac{\delta(\omega-\omega_0)}{\frac{\partial(\omega-E+E^{\prime})}{\partial \omega}}= \frac{\delta(\omega-\omega_0)}{1+\frac{\partial E^{\prime}}{\partial \omega}}\,.
\end{equation}
The partial derivative is easily obtained using \eq{eq:energy} and \eq{eq:momentum_conv},
\begin{equation} \label{eq:partial_E}
    \frac{\partial E^{\prime}}{\partial \omega} = \frac{\omega \cos^2\theta}{E^{\prime}-m^{\prime}\Omega}\,.
\end{equation}
We can now perform the $\omega$ integral in \eq{eq:W_expr_1} using \eq{eq:partial_E} and \eq{eq:delta_rewrite}, which gives us
\begin{equation} \label{eq:W_expr_2}
    W_{n,a} = \frac{q^2}{4\pi}\sum_{n^{\prime},a^{\prime}}\int d\theta  \frac{\omega_0^2\sin\theta(E'-m'\Omega)}{(E'-m'\Omega)+\omega_0\cos^2\theta} \left(\frac{1}{2} \sum_{\zeta \zeta^{\prime}} \Phi\right)\,.
\end{equation}
In \cite{Buzzegoli:2022dhw,Buzzegoli:2023vne}, we computed the radiation intensity numerically by evaluating the sum over $n'$ and $a'$ in \eq{eq:W_expr_2}. We will refer to that calculation as \emph{exact}. The scope of this work is to compute the  quasi-classical approximation for the radiation intensity. The applicability of the quasi-classical approximation requires that the main contribution to the intensity comes about from the large values of the quantum numbers $n\gg 1$ and $m\gg 1$, which allows treating them as continuous parameters. In particular, summations over quantum numbers can be replaced by integration. In the following two sections, we will expand the various expressions from this section in the quasi-classical limit to obtain the quasi-classical formula for the photon emission intensity.

\section{WKB Approximation of the \texorpdfstring{$I-$}{I-}Function}
\label{sec:wkb}

 In this section, we develop the quasi-classical approximation of the $I$-function. In the system under consideration, the motion of ultra-relativistic fermions is characterized by the large value of the principal quantum number $n$. Therefore, we are interested to obtain an approximation for $I_{n,n^{\prime}}(x)$ for large $n$ and $n'$. We follow the approach of Sokolov and Ternov in \cite{book:SokolovAndTernov}, who developed the WKB approximation of the $I$-function in the non-rotating case. Rotation requires significant modifications to their method as we proceed to discuss. 
 
First, consider the $I$-function in more detail in relation to the WKB procedure. The $I$-functions are solutions to the differential equation,
\begin{equation} \label{eq:I func diff eqn}
    \frac{d^2}{dx^2}(x^{1/2}y(x)) - f_{n,n'}(x)x^{1/2}y(x) =0\,,
\end{equation}
where 
\begin{equation} \label{eq:I func potential}
    f_{n,n'}(x)=\frac{(n-n')^2-1}{4 x^2} - \frac{(n+n'+1)}{2 x} + \frac{1}{4}\,.
\end{equation}
The differential equation is solved by $y(x)=I_{n,n'}(x)$, as defined in \eq{eq:I-func_def}. 

\begin{figure}[t]
    \includegraphics[width=4in]{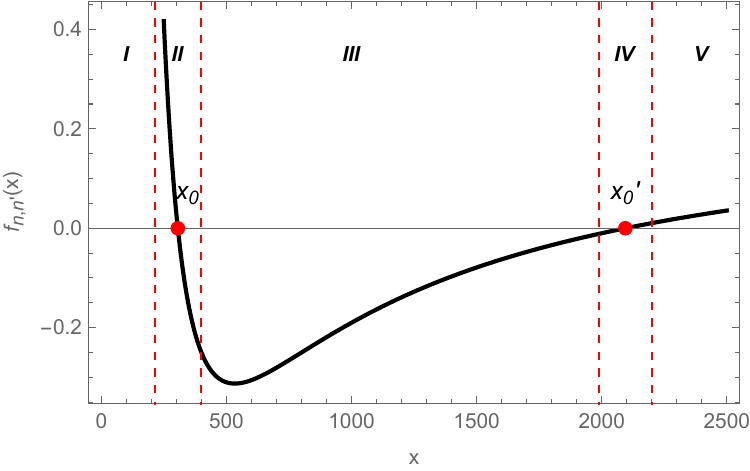}
  \caption{The function $f_{n,n'}(x)$ is plotted for $n=1000$, and $n^{\prime}=200$. The red points denote the two turning points (roots). The vertical lines separate the domain into five regions. Regions \text{I}, \text{III}, and \text{V}  are where WKB is applicable. In regions \text{II} and \text{IV}, the approximation is obtained by linearizing the potential. }
\label{fig:potential}
\end{figure}

Note that mathematically, Eq.~\eq{eq:I func diff eqn} is the Schrödinger equation with $x^{1/2}y(x)$ being the wave function, $f_{n,n'}(x)$, the potential, and $0$ as the energy eigenvalue. The turning points of the potential are the solutions to $f_{n,n'}(x)=0$ since ``energy" is zero. The ``potential" is plotted in \fig{fig:potential} for representative values of $n$ and $n'$. Eq.~\eq{eq:I func potential} implies that the turning points (for large $n$ and $n'$) are
\begin{align}\label{eq:turning_pt}
    x_0 = (\sqrt{n}-\sqrt{n'})^2\,,\qquad 
    x^{\prime}_0 = (\sqrt{n}+\sqrt{n'})^2\,.
\end{align}

When $\Omega=0$, the kinematics of the problem restricts the argument of the $I$ function, $x$, to be in close proximity and on the left of the first turning point $x_0$: $x<x_0$. An approximation that is valid in this region (see Fig.\ref{fig:potential}) is the relevant interpolation function used in the WKB procedure by linearizing the potential near its turning points. Such interpolation functions are proportional to Airy functions, but since in the non-rotating case, the argument $x$ is always to the left of the potential well, Airy functions can be represented by the modified Bessel functions $K_\nu$. For this reason, in \cite{book:SokolovAndTernov}, Sokolov and Ternov use $K_\nu$ functions as the approximation to the $I$-function in their quasi-classical calculations.

However, when rotation is present, $x$ can go beyond the turning point and into the potential well. For $I_{n,n'}(x)$, $x$ continues to remain close to the first turning point $x_0$. The approximation in \cite{book:SokolovAndTernov} would suffice to handle the case, except using the Airy function in place of the modified Bessel functions. However, the matrix element in Eq.~\eq{eq:matrix_element} has an additional $I$-function in the form of $I_{a,a'}(x)$, which is not restricted to the region near its corresponding turning point.  When $\Omega=0$, the above mentioned $I$-function does not appear because of the following property:
\begin{align} \label{eq:I func norm }
    \sum_{a'=0}^{\infty}I_{a,a'}(x)^2 = 1\,.
\end{align}
In the non-rotating case, the rest of the matrix element does not contain any dependence on $a'$ and hence the sum of $I_{a,a^{\prime}}(x)$ factors out, which does not happen when $\Omega \neq 0$. Therefore, we must develop the WKB approximation for the $I$-function that is valid for all $x>0$.

The WKB method, when applied to an equation of the form \eq{eq:I func diff eqn}, gives a piecewise solution in the three regions marked off by the two turning points. The $I$-function vanishes at $x=0$ and as $x \rightarrow \infty$, so we require that the WKB solution also satisfy the same conditions. Furthermore, the quasi-classical wave function is valid everywhere except in the close vicinity of the turning points $x_0$ and $x_0'$ where one has to employ the exact solution to the differential equation \eq{eq:I func diff eqn} with the linearized potential. This is also essential for numerical computations because WKB solutions diverge at the turning points. We employ the matching conditions for the five pieces of function to obtain the following form for $x^{1/2}y(x)$,
\begin{align} \label{eq:WKB_formula}
    x^{1/2} y(x) = N\begin{cases}
      \frac{1}{2\sqrt{\pi}} \frac{|f^{\prime}_{n,n'}(x_0)|^{1/6}}{[f_{n,n'}(x)]^{1/4}} \exp\left\{-\int_{x}^{x_0}\sqrt{f_{n,n'}(x^{\prime})}dx^{\prime}\right\} 
    & x\leq  x_0-\delta_1\, ,\\ 
       \Ai\left(|f^{\prime}_{n,n'}(x_0)|^{1/3}(x_0-x)\right)
& x_0-\delta_1< x<  x_0+\delta_2\,, \\
     \frac{1}{\sqrt{\pi}} \frac{|f^{\prime}_{n,n'}(x_0)|^{1/6}}{[-f_{n,n'}(x)]^{1/4}} \sin \left(\int_{x_0}^{x}\sqrt{(-f_{n,n'}(x'))}dx'+\frac{\pi}{4}\right)
& x_0+\delta_2\leq x\leq  x_0^{\prime}-\delta_3\,, \\ 
       (-1)^{p+1}\left(\frac{|f^{\prime}_{n,n'}(x_0)|}{f^{\prime}_{n,n'}(x_0^{\prime})}\right)^{1/6} \Ai\left(|f^{\prime}_{n,n'}(x_0)|^{1/3}(x-x_0^{\prime})\right)\,
       & x_0^{\prime}-\delta_3< x<  x_0^{\prime}+\delta_4\,, \\ 
     (-1)^{p+1}\frac{1}{2\sqrt{\pi}} \frac{|f^{\prime}_{n,n'}(x_0)|^{1/6}}{[-f_{n,n'}(x)]^{1/4}} \exp\left\{-\int_{x_0^{\prime}}^{x}\sqrt{f_{n,n'}(x^{\prime})}dx^{\prime}\right\}\, 
       &  x_0^{\prime}+\delta_4 \leq x\,. \\ 
   \end{cases}
\end{align}
Here $N$ is the normalization constant, $\delta_1,\ldots,\delta_4$  mark the left and right vicinities of the two turning points, and $p$ is the integer given by the Bohr-Sommerfeld quantization condition:
\begin{equation} \label{eq:quantization}
        \int_{x_0}^{x_0^{\prime}}\sqrt{(-f_{n,n'}(x'))}dx' = \left(p-\frac{1}{2}\right)\pi\,.
\end{equation}
The integral on the left-hand side of \eq{eq:quantization} can be evaluated exactly, but we need only the large $n$ and $n'$ limit where 
\begin{equation} \label{eq:potential}
    f_{n,n'}(x) \approx \frac{(x-x_0)(x-x_0^{\prime})}{4x^2}\,.
\end{equation}
Integration in \eq{eq:quantization} yields 
\begin{equation} \label{eq:quantization_2}
    \int_{x_0}^{x_0^{\prime}}\sqrt{(-f_{n,n'}(x'))}dx' =\pi \left( \min \{n,n^{\prime}\}+\frac{1}{2}\right)\,.
\end{equation}
This implies that 
\begin{equation} \label{eq:p-value}
    p = \min \{n,n^{\prime}\}+1\,.
\end{equation}
 With the same accuracy, the derivative of the function $f_{n,n'}$ reads
\begin{equation} \label{eq:potential_derivative}
    f^{\prime}_{n,n'}(x_0) \approx -\frac{\sqrt{n n^{\prime}}}{x_0^2}\,.
\end{equation}
Furthermore, the integrals that appear in \eq{eq:WKB_formula} can be evaluated as follows:
\begin{align}
S_1 =& \int_{x}^{x_0}\sqrt{f_{n,n'}(x^{\prime})}dx^{\prime}\\
  =&-\frac { 1 } { 2 }  \left[\sqrt{\left(x-x_0\right)\left(x-x_0^{\prime}\right)}+\sqrt{x_0 x_0^{\prime}} \ln \left(\frac{x+\sqrt{x_0 x_0^{\prime}}-\sqrt{\left(x-x_0\right)\left(x-x_0^{\prime}\right)}}{-x+\sqrt{x_0 x_0^{\prime}}+\sqrt{\left(x-x_0\right)\left(x-x_0^{\prime}\right)}}\right)\right. \nonumber\\
 &+\frac{\left(x_0+x_0^{\prime}\right)}{2} \ln \left|\frac{2}{\left(x_0^{\prime}-x_0\right)}\left(\left.x-\sqrt{\left(x-x_0\right)\left(x-x_0^{\prime}\right)}-\frac{\left(x_0+x_0^{\prime}\right)}{2}\right)\right| 
 -  \sqrt{x_0 x_0^{\prime}} \ln \left(\frac{x_0+\sqrt{x_0 x_0^{\prime}}}{-x_0+\sqrt{x_0 x_0^{\prime}}}\right)\right]\,,\label{eq:integral_1}\nonumber\\
S_2= & \int_{x_0}^{x}\sqrt{-f_{n,n'}(x^{\prime})}dx^{\prime}\\
= & \frac { 1 } { 2 } \left[\sqrt{\left(x-x_0\right)\left(x_0^{\prime}-x\right)}+2\sqrt{x_0 x_0^{\prime}}\tan^{-1}\left(\sqrt{\frac{x_0}{x_0^{\prime}}} \left(\frac{x_0^{\prime}-x}{x-x_0}\right)\right)\right.\nonumber\\ 
& \left. \hphantom{\sqrt{\left(x-x_0\right)\left(x_0^{\prime}-x\right)}+}
-(x_0+x_0^{\prime}) \tan^{-1}\left(\sqrt{\frac{x_0^{\prime}-x}{x-x_0}}\right)+\frac{\pi}{2}\left(x_0+x_0^{\prime}-2\sqrt{x_0 x_0^{\prime}}\right)\right]\,,
\label{eq:integral_2}\nonumber\\
S_3=&  \int_{x_0^{\prime}}^{x}\sqrt{f_{n,n'}(x^{\prime})}dx^{\prime}\\
=& \frac { 1 } { 2 } \left[\sqrt{\left(x-x_0\right)\left(x-x_0^{\prime}\right)}+\sqrt{x_0 x_0^{\prime}} \ln \left(\frac{x+\sqrt{x_0 x_0^{\prime}}-\sqrt{\left(x-x_0\right)\left(x-x_0^{\prime}\right)}}{-x+\sqrt{x_0 x_0^{\prime}}+\sqrt{\left(x-x_0\right)\left(x-x_0^{\prime}\right)}}\right)\right. \nonumber\\
& +\frac{\left(x_0+x_0^{\prime}\right)}{2} \ln \left|\frac{2}{\left(x_0^{\prime}-x_0\right)}\left(\left.x-\sqrt{\left(x-x_0\right)\left(x-x_0^{\prime}\right)}-\frac{\left(x_0+x_0^{\prime}\right)}{2}\right)\right|
-\sqrt{x_0 x_0^{\prime}} \ln \left(\frac{x_0^{\prime}+\sqrt{x_0 x_0^{\prime}}}{-x_0^{\prime}+\sqrt{x_0 x_0^{\prime}}}\right)\right]\,. \nonumber
\label{eq:integral_3}
\end{align}

We can fix the overall normalization constant $N$ by employing the approximation used by Sokolov and Ternov in \cite{sokolov1986radiation} for the $I$-function near the first turning point, $x_0$, as $x \rightarrow x_0$ from the left:
\begin{equation} \label{eq:Sokolov_Ternov_approx}
    I_{n,n^{\prime}}(x) \approx \frac{1}{\pi \sqrt{3}}\left(1-\frac{x}{x_0}\right)^{1/2}K_{1/3}(z) \,,
\end{equation}
where $K_{1/3}(z)$ is the Bessel-$K$ function and $z$ is defined as
\begin{equation} \label{eq:z-variable}
    z = \frac{2}{3}(x_0^2 n n^{\prime})^{1/4}\left(1-\frac{x}{x_0}\right)^{3/2}\,.
\end{equation}
 In \eq{eq:WKB_formula}, $y(x)$ near the turning point $x_0$ is an Airy function, which can be expressed in terms of the modified Bessel function:
 \begin{equation}
     \Ai(u) =\frac{1}{\pi \sqrt{3}} \sqrt{u}\, K_{1/3}\left(\frac{2}{3}u^{3/2}\right)\,,\quad   u>0\,.
 \end{equation}
 Therefore, we can express $y(x)$ near the turning point $x_0$ as follows:
 \begin{equation} \label{eq:normalization_matching}
     y(x) = N\left(\frac{x_0}{x}\right)^{1/2}\frac{1}{\pi\sqrt{3}}\frac{(n n^{\prime})^{1/12}}{x_0^{1/3}}\left(1-\frac{x}{x_0}\right)^{1/2}K_{1/3}\left(\frac{2}{3}(x_0^2 n n^{\prime})^{1/4}\left(1-\frac{x}{x_0}\right)^{3/2}\right)\,.
 \end{equation}
 Comparing \eq{eq:normalization_matching} with \eq{eq:Sokolov_Ternov_approx} and noting that $x \approx x_0$ we can fix the normalization constant 
 \begin{equation} \label{eq:normalization}
     N = \frac{x_0^{1/3}}{(n n^{\prime})^{1/12}}\,,
 \end{equation}
using which we obtain the desired approximation to the $I$-function:
\begin{align} \label{eq:I_func_approx}
    I_{n,n'}(x) = \begin{cases}
      \frac{1}{\sqrt{2\pi}} \frac{1}{(x_0-x)^{1/4}(x_0^{\prime}-x)^{1/4}} \exp\left\{-S_1\right\} 
       & x\leq  x_0-\delta_1\,, \\
 \frac{x_0^{1/3}}{(n n^{\prime})^{1/12}}\frac{1}{x^{1/2}} \Ai\left(\frac{(n n^{\prime})^{1/6}}{x_0^{2/3}}(x_0-x)\right)
       & x_0-\delta_1< x<  x_0+\delta_2\,, \\ 
      \sqrt{\frac{2}{\pi}} \frac{1}{(x-x_0)^{1/4}(x_0^{\prime}-x)^{1/4}}  \sin \left(S_2+\frac{\pi}{4}\right)\
       & x_0+\delta_2\leq x\leq  x_0^{\prime}-\delta_3\,, \\
      (-1)^{p+1} \frac{x_0^{\prime 1/3}}{(n n^{\prime})^{1/12}}\frac{1}{x^{1/2}}
      \Ai\left(\frac{(n n^{\prime})^{1/6}}{x_0^{\prime 2/3}}(x-x_0^{\prime})\right)
       & x_0^{\prime}-\delta_3< x<  x_0^{\prime}+\delta_4\,, \\
      (-1)^{p+1} \frac{1}{\sqrt{2\pi}} \frac{1}{(x-x_0)^{1/4}(x-x_0^{\prime})^{1/4}}  \exp\left\{-S_3\right\} 
       &  x_0^{\prime}+\delta_4 \leq x\,. \\ 
   \end{cases}
\end{align}
The parameters $\delta_\alpha$, $\alpha=1,\ldots,4$ allow a continuous interpolation between the five different regions shown in \fig{fig:potential}. Since the WKB approximation breaks down at distances shorter than the de Broglie wavelength $\lambdabar$ from the turning points,  $\delta_\alpha$'s must be larger than $\lambdabar$. Noting that $\sqrt{-f}$ is the  ``momentum" in Schr\"odinger's equation \eq{eq:I func diff eqn}, $\delta_\alpha$ must satisfy the condition 
\begin{align}\label{deltas}
\delta_\alpha\gtrsim \frac{1}{\sqrt{-f}}\,.
\end{align}
Expanding $f(x)\approx f'(x_0)(x-x_0)$ near the left turning point and using \eq{eq:potential_derivative} and \eq{eq:turning_pt} we obtain the condition 
\begin{align}\label{deltas2}
\delta_1,\delta_2\gtrsim\frac{x_0}{(nn')^{1/6}(\sqrt{n}-\sqrt{n'})^{1/3}}\,.
\end{align}
A similar condition holds for the right turning point:
\begin{align}\label{deltas3}
\delta_3,\delta_4\gtrsim\frac{x_0'}{(nn')^{1/6}(\sqrt{n}+\sqrt{n'})^{1/3}}\,.
\end{align}
Any choice of $\delta_\alpha$ satisfying \eq{deltas2}, \eq{deltas3} is within the WKB accuracy. 
In \sec{sec:results}, we discuss the values of $\delta_\alpha$ chosen for  the numerical calculation.

Eq.~\eq{eq:I_func_approx} is our main result in this section. The substitution $n\to a$, $n'\to a'$ yields $I_{a,a'}(x)$. The functions $I_{n,n'}(x)$ and $I_{a,a'}(x)$ are then plugged into the matrix element \eq{eq:matrix_element}. We note that although they have the same argument $x$, their turning points are different. As a result, whereas the argument of $I_{a,a'}(x)$ can take any positive value in relation to its turning points, the argument of  $I_{n,n'}(x)$ is always restricted to the vicinity of its left turning point $x_0$, viz.\ the second region in \eq{eq:I_func_approx}.

\section{Quasi-Classic approximation of Intensity}
\label{sec:quasiclassics}

We are now all set to derive the quasi-classical approximation for the radiation intensity. In addition to using the expressions for $I$-functions we derived in the previous section, the matrix elements and the phase space in Eq.~\eq{eq:intensity_formula} must be expanded in the ultra-relativistic limit  
\begin{align}
    E , E',\omega \gg  M, p_z\,.
\end{align}
We also assume that the rotation is slow enough that the following inequality holds:
\begin{align}
      | m\Omega|, |m'\Omega| \ll E , E^\prime ,\omega\,.
\end{align}

Expanding the coefficients in \eq{eq:coeff-2} to the lowest order in the small parameters gives us
\begin{subequations} \label{eq:coeff-1 approx}
    \begin{align}\label{eq:Csapprox}
     C_{1,2} &= \frac{1}{2\sqrt{2}}(1\pm\zeta)\left(1+\frac{1}{2}\frac{M}{E}\right)\,,\qquad
    C_{3,4} = \frac{1}{2\sqrt{2}}(1\mp\zeta)\left(1-\frac{1}{2}\frac{M}{E}\right)\,,\\
\label{eq:Csprime1approx}
     C_{1,2}' &= \frac{1}{2\sqrt{2}}\left[(1\pm\zeta')\left(1+\frac{1}{2}\frac{M}{E'}\right)\mp(1\mp\zeta')\frac{1}{2}\frac{\omega \cos{\theta}}{E'}\right]\,,\\
    \label{eq:Csprime2approx}
     C_{3,4}' & = \frac{1}{2\sqrt{2}}\left[(1\mp\zeta')\left(1-\frac{1}{2}\frac{M}{E'}\right)\mp(1\pm\zeta')\frac{1}{2}\frac{\omega \cos{\theta}}{E'}\right]\,.
    \end{align}
 \end{subequations}
 Plugging the expressions in \eq{eq:coeff-1 approx} into the equations in \eq{eq:Ks}, we have
\begin{subequations} \label{eq:K approx}
    \begin{align}\label{eq:K1 approx}    
        K_{1,2} &= \frac{1}{4}\left[(1+\zeta \zeta') \pm\zeta(1+\zeta\zeta') \frac{M \omega}{2E E'}\mp(1-\zeta \zeta')\frac{\omega \cos{\theta}}{2 E'}\right]\,,\\
 \label{eq:K2 approx}
     K_{3,4} &= \frac{1}{4}\left[(1-\zeta \zeta') \pm\zeta(1-\zeta\zeta') \frac{M \omega}{2EE'}\pm(1+\zeta \zeta')\frac{\omega \cos{\theta}}{2 E'}\right]\,.
    \end{align}
 \end{subequations}
 The matrix element in \eq{eq:matrix_elt_helcty_sum} contains several $I$-functions. They obey the recursion relations 
\begin{subequations} \label{eq:I_func_prop}
\begin{align}
    I_{n,n'-1}(x)&= \left(\frac{x}{n'}\right)^{1/2}\left[\left(\frac{n-n'-x}{2x}\right)I_{n,n'}(x)-I'_{n,n'}(x) \right]\,,\\
    I_{n-1,n'}(x)&= \left(\frac{x}{n}\right)^{1/2}\left[\left(\frac{n-n'+x}{2x}\right)I_{n,n'}(x)+I'_{n,n'}(x) \right]\,,\\
    I_{n-1,n'-1}(x)&= \frac{x}{(n n')^{1/2}}\left[\left(\frac{n+n'-x}{2x}\right)I_{n,n'}(x)-I'_{n,n'}(x) \right]\,,    
\end{align}
\end{subequations} 
where $I_{n,n'}$ and $I'_{n,n'}$ in the right-hand-side are replaced by their quasiclassical approximation \eq{eq:Sokolov_Ternov_approx}. However, since $\Omega \neq 0$ implies that the argument of the $I$-function can also be greater than the turning point $x_0$, we must convert the Bessel functions to Airy functions, which yields
\begin{subequations} \label{eq:I conversion}
    \begin{align} \label{eq:I to Airy}
    I_{n,n'}(x)& = \frac{1}{(x_0^2nn')^{1/12}} \Ai\left((x_0^2nn')^{1/6}\left(1-\frac{x}{x_0}\right)\right)\,,\\
\label{eq:Iprime to Airyprime}
    I'_{n,n'}(x)& = -\frac{(nn')^{1/12}}{x_0^{5/6}} \Ai'\left((x_0^2nn')^{1/6}\left(1-\frac{x}{x_0}\right)\right)\,.
\end{align}
\end{subequations}
The argument of the $I$-functions contains the ratio
\begin{align} \label{eq:ratio}
    \frac{x}{x_0} = \frac{\omega^2-\omega^2\cos^2{\theta}}{\left(\sqrt{(E-m\Omega)^2-M^2}-\sqrt{(E'-m'\Omega)^2-M^2-p_z^2}\right)^2}\,.
\end{align}
Expanding \eq{eq:ratio} to the lowest  order we get 
\begin{align} \label{eq:z calculation}
   (x_0^2n n')^{1/6} \left(1-\frac{x}{x_0}\right) =  \left(\frac{\omega \sqrt{E E'}}{2 |q B|}\right)^{2/3}\Bigg[\left(\frac{E}{E'}\right)\left(\frac{p_z^2}{\omega^2}+\frac{M^2}{E^2}\right)-\frac{\Omega}{3}\left(\frac{8(m-m')}{\omega}
   +\frac{m}{E}+\frac{m'}{E'}\right) \Bigg] \,.
\end{align}
 For convenience of notation, we will define two new parameters $\varepsilon$ and $\kappa$:
\begin{align} \label{eq:epsilon}
    &\varepsilon =  \left(\frac{p_z^2}{\omega^2}+\frac{M^2}{E^2}\right)-\left(\frac{E'}{E}\right)\Bigg[\frac{\Omega}{3}\left(\frac{8(m-m')}{\omega}
   +\frac{m}{E}+\frac{m'}{E'}\right) \Bigg] \,,\\
    &\kappa = \frac{\omega E^2}{2|qB| E'}\,.\label{eq:epsilon-2}
\end{align}
Collecting \eq{eq:I_func_prop}, \eq{eq:I conversion}, \eq{eq:z calculation} and using Eqs. \eq{eq:epsilon}, \eq{eq:epsilon-2} we derive the following approximate expressions for various $I$-functions:
\begin{subequations} \label{eq:I func QC}

\begin{align}
I_{n,n'}(x) =& \frac{1}{\kappa^{1/3}}\left(\frac{E}{E'}\right)^{1/2} \Bigg\{1+\frac{\Omega}{6}\Bigg[\frac{2(m-m')}{\omega}+\frac{m}{E}+\frac{m'}{E'}\Bigg]\Bigg\} \Ai\left(\kappa^{2/3}\varepsilon\right), \label{eq:I func QC 1}\\
I'_{n,n'}(x) =& -\frac{1}{\kappa^{2/3}}\frac{E^{3/2}}{\omega E^{1/2}} \Bigg\{1+\frac{\Omega}{3}\Bigg[\frac{5(m-m')}{\omega}-\frac{m}{2E}-\frac{m'}{2E'}\Bigg]\Bigg\} \Ai'\left(\kappa^{2/3}\varepsilon\right), \label{eq:I func QC 2}\\
I_{n,n'-1}(x) =& \frac{1}{\kappa^{1/3}}\left(\frac{E}{E'}\right)^{1/2}\Bigg\{\left(1+\Omega\Bigg[\frac{(m-m')}{\omega}\left(\frac{1}{3}-\frac{E}{E'}\right)+\frac{m}{6E}+\frac{m'}{6E'}\Bigg]\right)\Ai\left(\kappa^{2/3}\varepsilon\right)\nonumber \\
     &+\frac{1}{\kappa^{1/3}}\left(\frac{E}{E'}\right)\left(1+\frac{\Omega}{3}\Bigg[\frac{5(m-m')}{\omega}+\frac{5m'}{2E'}-\frac{m}{2E}\Bigg]\right)\Ai'\left(\kappa^{2/3}\varepsilon\right)\Bigg\}, \label{eq:I func QC 3}\\
I_{n-1,n'}(x) =&  \frac{1}{\kappa^{1/3}}\left(\frac{E}{E'}\right)^{1/2}\Bigg\{\left(1+\Omega\Bigg[\frac{(m-m')}{\omega}\left(\frac{1}{3}-\frac{E'}{E}\right)+\frac{m}{6E}+\frac{m'}{6E'}\Bigg]\right)\Ai\left(\kappa^{2/3}\varepsilon\right)\nonumber \\
        &-\frac{1}{\kappa^{1/3}}\left(1+\frac{\Omega}{3}\Bigg[\frac{5(m-m')}{\omega}+\frac{5m}{2E}-\frac{m'}{2E'}\Bigg]\right)\Ai'\left(\kappa^{2/3}\varepsilon\right)\Bigg\} \label{eq:I func QC 4},\\
I_{n-1,n'-1}(x) =& \frac{1}{\kappa^{1/3}}\left(\frac{E}{E'}\right)^{1/2}\Bigg\{\left(1+\Omega\Bigg[\frac{(m-m')}{\omega}\left(\frac{1}{3}-\frac{\omega^2}{EE'}\right)+\frac{m}{6E}+\frac{m'}{6E'}\Bigg]\right)\Ai\left(\kappa^{2/3}\varepsilon\right)\nonumber \\
        &+\frac{1}{\kappa^{1/3}}\frac{\omega}{E'}\left(1+\frac{5\Omega}{3}\Bigg[\frac{(m-m')}{\omega}+\frac{m}{2E}+\frac{m'}{2E'}\Bigg]\right)\Ai'\left(\kappa^{2/3}\varepsilon\right)\Bigg\}.\label{eq:I func QC 5}
    \end{align}
\end{subequations}
Combining these expressions with the coefficients in Eq. \eq{eq:K approx} we obtain the matrix element $\Phi$, defined in Eq. \eq{eq:matrix_elt_helcty_sum}, up to the order $\mathcal{O}(M^2/ E^2, \Omega/E)$, that is
\begin{align}\label{eq:matrix_elt_expanded}
    \Phi =& \frac{I^2_{a,a'}(x)}{4\kappa^{2/3}}\left(\frac{E}{E'}\right)\Bigg\{\frac{(1+\zeta\zeta')}{2}\Bigg[\left(2+\frac{\omega}{E'}\right)^2\cos^2\theta\Ai^2(\kappa^{2/3}\varepsilon)+\Bigg(\frac{1}{\kappa^{1/3}}\left(2+\frac{\omega}{E'}\right)\Ai'(\kappa^{2/3}\varepsilon)\nonumber \\
    & +\Bigg(\frac{\zeta M\omega}{EE'}-\frac{(\omega+2E')}{EE'}(m-m')\Omega\Bigg)\Ai(\kappa^{2/3}\varepsilon)\Bigg)^2\Bigg]+\frac{(1-\zeta\zeta')}{2}\Bigg[\frac{\omega^2\cos^2\theta}{E'^2}\Ai^2(\kappa^{2/3}\varepsilon)\nonumber \\
    &+\Bigg(\left(\frac{\zeta M\omega}{E E'}+\frac{\omega (m-m')\Omega}{E E'}\right)\Ai(\kappa^{2/3}\varepsilon)-\frac{\omega}{\kappa^{1/3}E'}\Ai'(\kappa^{2/3}\varepsilon)\Bigg)^2\Bigg]\Bigg\}\,.
\end{align}
Summing over $\zeta^{\prime}$ and averaging over $\zeta$, we have
\begin{align}\label{eq:matrix_elt_final}
     \frac{1}{2}\sum_{\zeta =\pm 1}\sum_{\zeta' =\pm 1} \Phi =& \frac{I^2_{a,a'}(x)}{4\kappa^{2/3}}\left(\frac{E}{E'}\right)\Bigg\{\Bigg[\frac{2M^2\omega^2}{E^2E'^2}+\Bigg(\left(2+\frac{\omega}{E'}\right)^2+\frac{\omega^2}{E'^2}\Bigg)\frac{(m-m')^2\Omega^2}{E^2}\Bigg]\Ai^2(\kappa^{2/3}\varepsilon)\nonumber \\
    &+\Bigg[\left(2+\frac{\omega}{E'}\right)^2+\frac{\omega^2}{E'^2}\Bigg]\cos^2\theta\Ai^2(\kappa^{2/3}\varepsilon)+\Bigg[\left(2+\frac{\omega}{E'}\right)^2+\frac{\omega^2}{E'^2}\Bigg]\frac{1}{\kappa^{2/3}}\Ai'^2(\kappa^{2/3}\varepsilon)\nonumber \\
    &-2\Bigg[\left(2+\frac{\omega}{E'}\right)^2+\frac{\omega^2}{E'^2}\Bigg]\frac{(m-m')\Omega}{E}\frac{1}{\kappa^{1/3}}\Ai(\kappa^{2/3}\varepsilon)\Ai'(\kappa^{2/3}\varepsilon)\Bigg\}\,.
\end{align}

The crux of the quasi-classical approximation is that at high energies, the discrete energy levels can be treated as continuous. The assumptions of the quasi-classical approximation, therefore, enable us to replace the summations over $n^{\prime}$ and $a^{\prime}$ in Eq. (\ref{eq:W_expr_2}) with the corresponding integrals, treating these quantum numbers as continuous variables. We thus obtain the quasi-classical approximation of the total radiation intensity:
\begin{equation} \label{W_expr_final}
    W_{n,a} = \frac{q^2}{4\pi}\int_{0}^{n} d n^{\prime} \int_{0}^{\infty} d a^{\prime} \int_{0}^{\pi} d\theta  \frac{\omega_0^2\sin\theta(E'-m'\Omega)}{(E'-m'\Omega)+\omega_0\cos^2\theta}\left( \frac{1}{2} \sum_{\zeta \zeta^{\prime}} \Phi\right)\,,
\end{equation}
with the summation over $\Phi$ given by Eq. (\ref{eq:matrix_elt_final}).  This is our main result.

\section{Numerical results}
\label{sec:results}

\begin{figure}[ht]
    \begin{tabular}{cc}
      \includegraphics[width=0.48\linewidth]{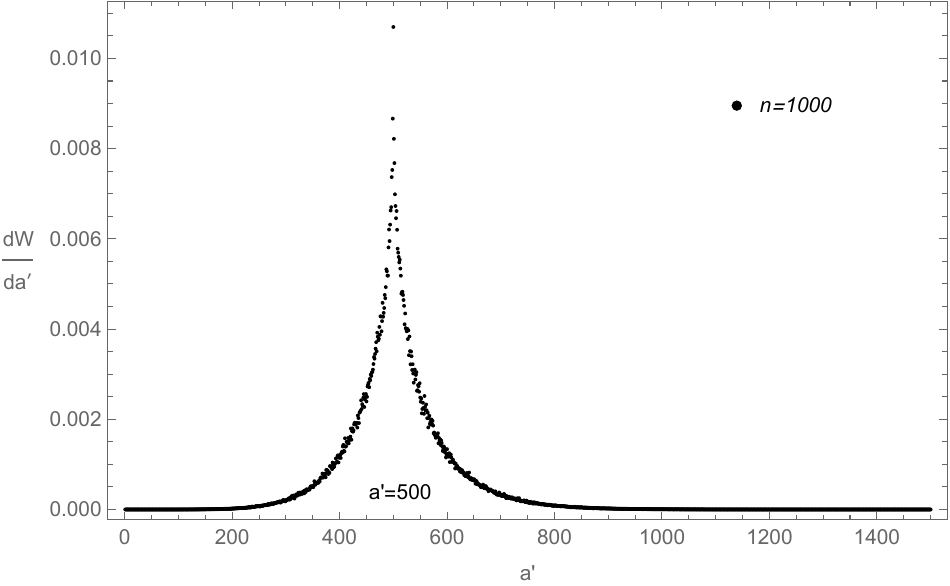}&
      \includegraphics[width=0.48\linewidth]{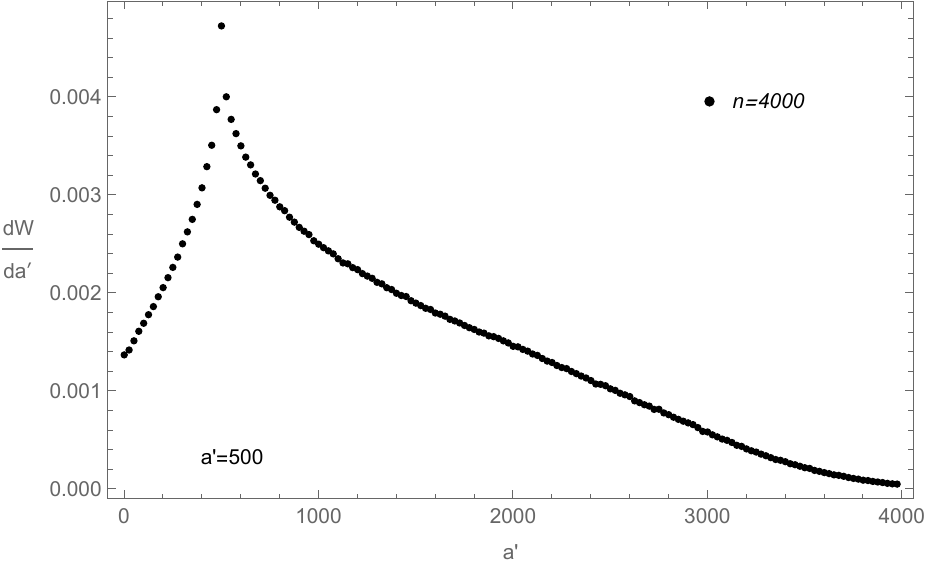} \\
      \includegraphics[width=0.48\linewidth]{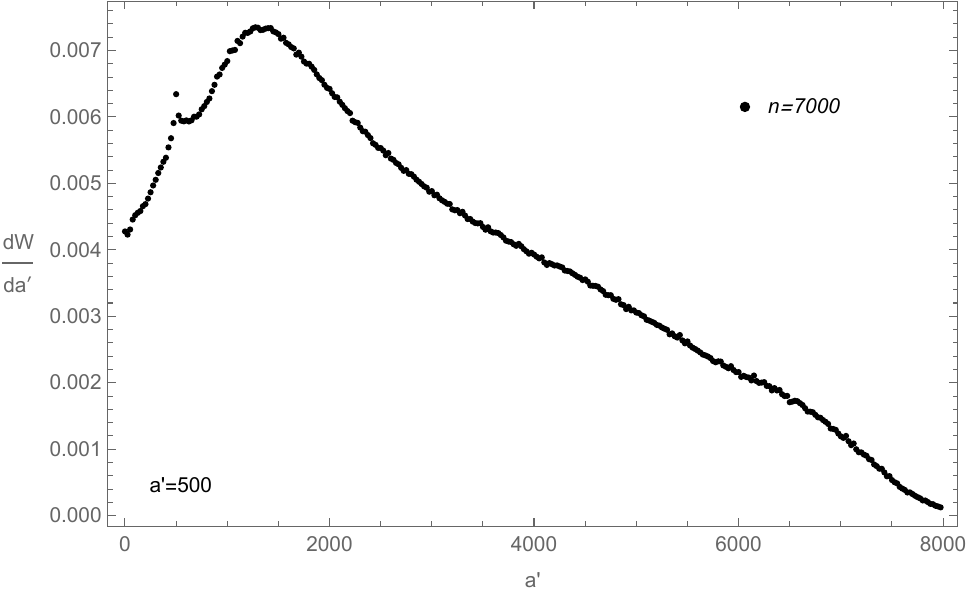}&
      \includegraphics[width=0.48\linewidth]{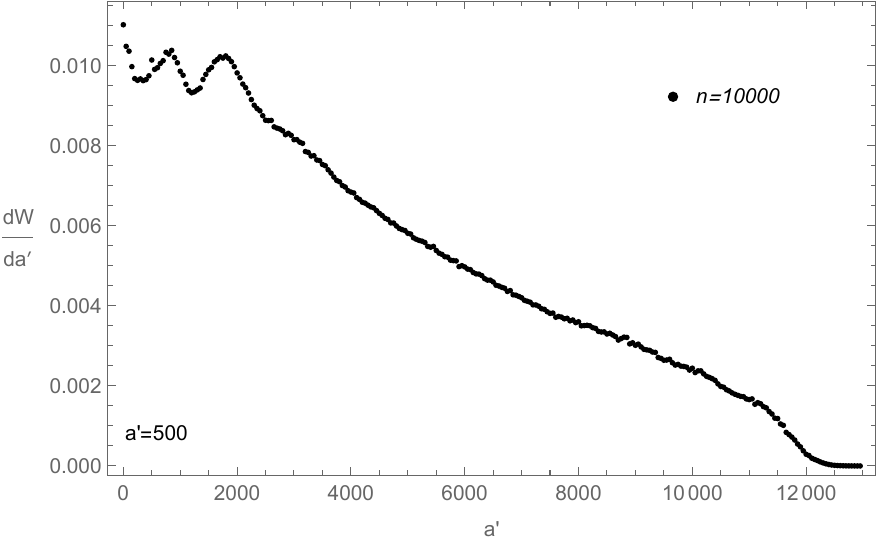} \\
    \end{tabular}
  \caption{The differential radiation intensity in units of $M^2$ for $n = 1000$, $4000$, $7000$ and $10000$, for $\Omega = 10^{-5}$, $a = 500$, and $qB = -0.01$.}
\label{fig:differential}
\end{figure}

To compute the radiation intensity, we  first investigated the differential intensity as a function of the radial quantum numbers $a,\, a'$. The parameters $\delta_\alpha$ are fixed to be $\delta_1=\delta_2 = x_0/10$ and $\delta_3=\delta_4 = x_0'/10$ which satisfy the requirements \eq{deltas2} and \eq{deltas3} for $n,\,n'>100$. We verified that our results do not depend on the choice of these parameters so long as they satisfy the WKB condition, i.e., they are close enough to their turning points. A few representative spectra are shown in \fig{fig:differential}. The key observation is that the initial and final quantum numbers $a$ and $a'$ are strongly correlated. This correlation becomes less pronounced at higher energies/larger $n$. Nevertheless, it is instrumental in controlling the accuracy of the $a'$-integration in Eq. \eq{W_expr_final}, which we carry out using a standard Monte Carlo algorithm.

\begin{figure}[ht]
      \includegraphics[width=0.6\linewidth]{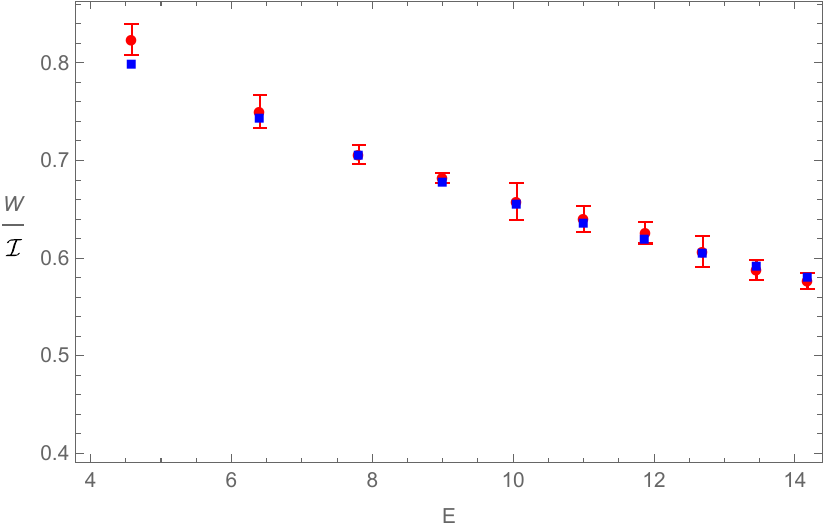}
  \caption{Validation of our numerical code in the stationary case $\Omega=0$ for the total radiation intensity $W$ as a function of fermion energy $E$ at  $a = 500$, $M=1$, and $qB = -0.01$. Red circles and error bars represent  Eq.~\eq{W_expr_final} computed with the Monte Carlo method. Blue squares represent the well-known analytical expression given by Eq.~\eq{eq:ST} \cite{berestetskii1982quantum}.  }
\label{fig:zero_rotation}
\end{figure}
\begin{figure}[t]
\begin{tabular}{cc}
      \includegraphics[width=3.4in]{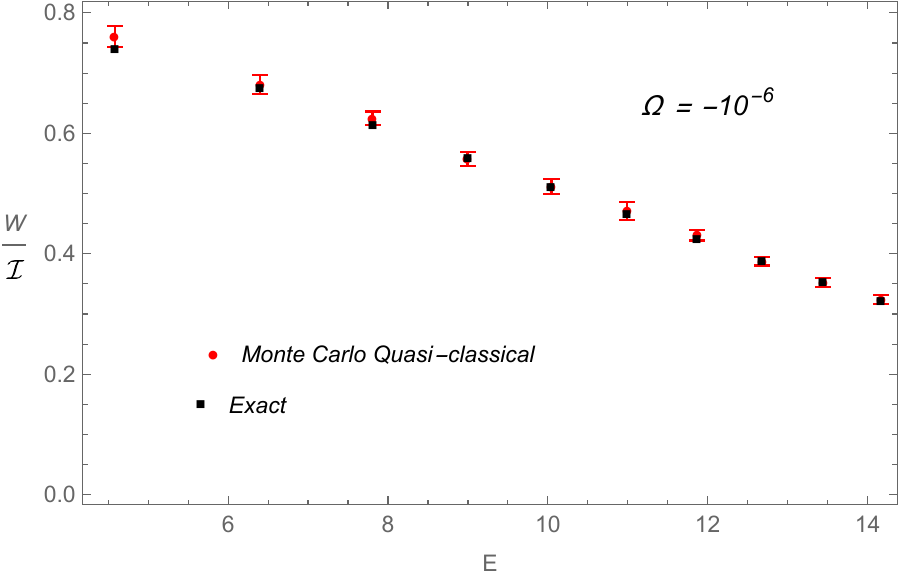}&
      \includegraphics[width=3.4in]{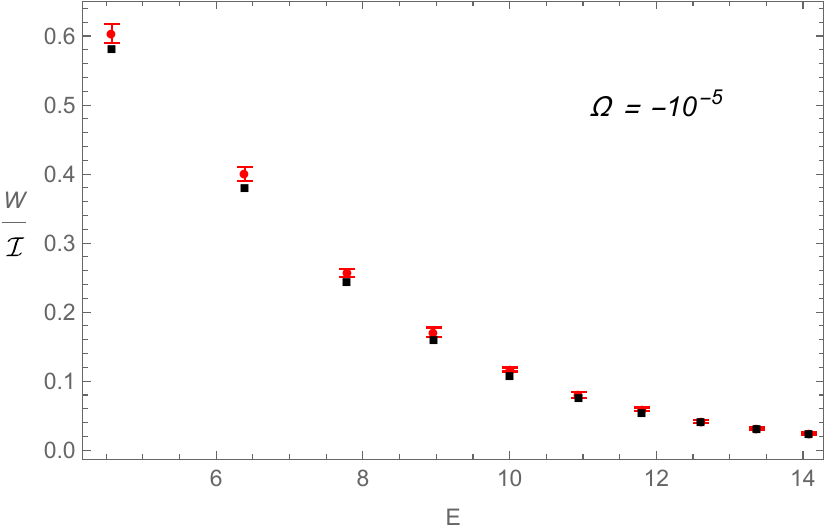} \\
      \includegraphics[width=3.4in]{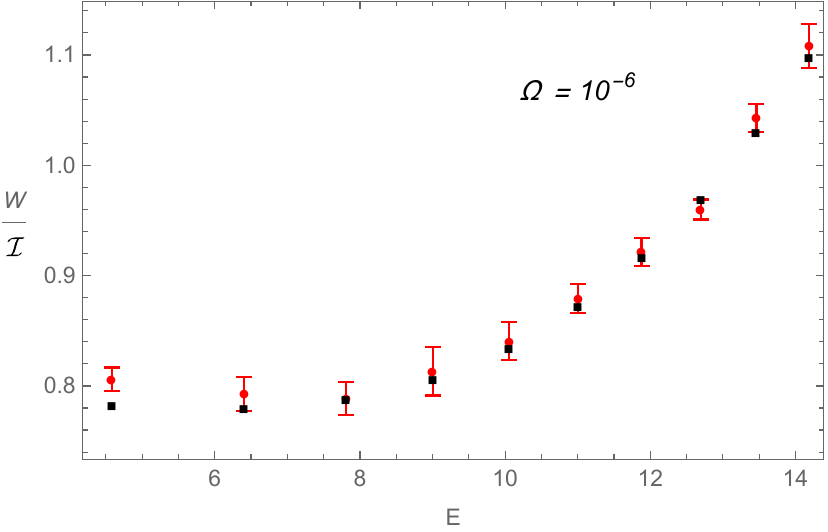}&
      \includegraphics[width=3.4in]{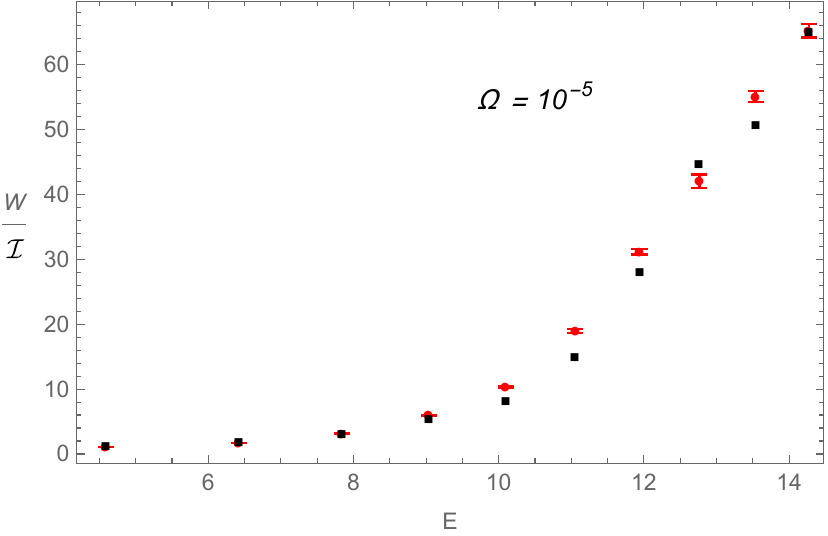} \\
\end{tabular}
  \caption{The intensity of photon emission is compared for different values of angular velocity $\Omega$ as a function of the initial fermion energy $E$ for a negative charge. We set $M=1$, and $B=0.01$. The red circles are the intensities computed from the quasi-classical formula derived in this paper, while the black squares denote the intensities from the exact computations performed in \cite{buzzegoli2023photon} for the same input parameters. The top figures are for $\Omega<0$ (meaning that $\bm B$ and $\bm\Omega$ are antiparallel) and the bottom figures are for $\Omega>0$ ($\bm B$ and $\bm\Omega$ are parallel). The figures on the left are when $|\Omega|=10^{-6}$ and the figures on the right are when $|\Omega|=10^{-5}$. 
  }
\label{fig:IntensityResults}
\end{figure}

In the figures below, we plot the radiation intensity in units of 
\begin{align}
\mathcal{I}= \frac{q^2}{4\pi} \frac{2}{3} \frac{(qBE)^2}{M^4}\,,
\end{align}
which represents the high energy limit of the classical radiation intensity $W_\text{cl}$. More precisely:
\begin{equation}
W_\text{cl} = \mathcal{I}\left(1-\frac{M^2}{E^2}\right)\,.
\end{equation}

To validate our numerical procedure, we computed the synchrotron intensity in the stationary limit, $\Omega= 0$, and verified that it agrees with the well-known result first obtained by Sokolov and Ternov \cite{sokolov1986radiation,berestetskii1982quantum}:
\begin{align}\label{eq:ST}
    \frac{W_\text{ST}}{\mathcal{I}} = -\frac{3}{4}\int_0^{\infty}\frac{4+5\chi x^{3/2}+4\chi^2x^3}{(1+\chi x^{3/2})^4}\Ai'(x)x dx\,,  
\end{align}
where 
\begin{align}
\chi = \frac{qBE}{M^3}\sqrt{1-\frac{M^2}{E^2}}\,,
\end{align}
as shown in  \fig{fig:zero_rotation}. It is seen in  \fig{fig:zero_rotation} that our Monte Carlo algorithm agrees with the analytical result with $\sim 5\%$ accuracy.

The results for the radiation intensity at finite $\Omega$ are exhibited in \fig{fig:IntensityResults} for different values and directions of rotation. They are compared with the calculations performed in \cite{buzzegoli2023photon} of the exact results in Eq. (\ref{eq:W_expr_2}).
We observe that they are generally in a good agreement. This indicates that our quasi-classical approximation provides an adequate description of the photon production at high energies and sub-critical magnetic fields. However, the numerical cost of the quasi-classical calculation is roughly two orders of magnitude less than with the exact expressions. We believe that the efficiency of the quasi-classical algorithm can be further significantly improved.

\section{Summary and outlook}\label{sec:summary}

We developed the quasi-classical approximation for the synchrotron radiation of a fermion embedded in the rigidly rotating system and subject to the constant magnetic field. It is valid for ultra-relativistic fermions under the conditions explained in the text. We numerically verified that our semi-classical expressions accurately approximate the exact formulas. At the same time, they decrease the numerical cost of computations by at least several orders of magnitude.

In our analysis, we derived the quasi-classical approximation by expanding the exact leading order expressions. A different approach, developed by Baier and Katkov, is to compute the quasi-classical matrix element from the get-go by observing that in the ultra-relativistic limit, the photon is emitted from a short part of the classical fermion trajectory in the magnetic field \cite{berestetskii1982quantum,BKS}. It would be interesting to generalize this method to the case of rotating fermions.

In this work we focused on the dynamics of the synchrotron radiation by a single rotating fermion, but our ultimate goal is to compute the electromagnetic radiation by the quark-gluon plasma. Indeed, it was argued that its electromagnetic radiation contains a significant synchrotron radiation component \cite{Tuchin:2012mf,Yee:2013qma,Zakharov:2016kte,Wang:2020dsr,Wang:2020dsr}.   While the computation of the plasma electromagnetic radiation with account of the magnetic field and rotation with exact formulas requires enormous computing resources, the quasi-classical method offers a significantly more effective approach.

\acknowledgments

We thank J.D.~Kroth for providing the numerical values of intensity obtained with the exact formula (represented by the black points in \fig{fig:IntensityResults}) and for many informative discussions. This work was supported in part by the U.S.\ Department of Energy under Grant No.\ DE-SC0023692.
M.B. was supported by the European Union - NextGenerationEU through grant No.\ 760079/23.05.2023, funded by the Romanian Ministry of Research, Innovation and Digitization through Romania's National Recovery and Resilience Plan, call no. PNRR-III-C9-2022-I8.

\section*{Data Availability}

The data that support the findings of this article are not
publicly available. The data are available from the authors
upon reasonable request.

\bibliographystyle{apsrev4-1}
\bibliography{bibliography.bib}
\end{document}